\documentclass[prl,reprint,showpacs,groupedaddress]{revtex4-1}

\usepackage{graphicx}
\usepackage[colorlinks, citecolor=blue]{hyperref}
\begin{document}

\title{{\normalsize Temperature dependent three-dimensional anisotropy of the magnetoresistance in WTe$_2$}}

\author{L. R. Thoutam$^{1,2}$}
\author{Y. L. Wang$^{1}$}\email{ylwang@anl.gov}
\author{Z. L. Xiao$^{1,2}$}\email{xiao@anl.gov or zxiao@niu.edu}
\author{S. Das$^{3}$}
\author{A. Luican-Mayer$^{3}$}
\author{R. Divan$^{3}$}
\author{G. W. Crabtree$^{1,4}$}
\author{W. K. Kwok$^{1}$}

\affiliation{$^{1}$Materials Science Division, Argonne National Laboratory, Argonne, Illinois 60439, USA}

\affiliation{$^{2}$Department of Physics, Northern Illinois University, DeKalb, Illinois 60115, USA}

\affiliation{$^{3}$Center for Nanoscale Materials, Argonne National Laboratory, Argonne, Illinois 60439, USA}

\affiliation{$^{4}$Departments of Physics, Electrical and Mechanical Engineering, University of Illinois at Chicago, Illinois 60607, USA}

\date{\today}

\begin{abstract}
Extremely large magnetoresistance (XMR) was recently discovered in WTe$_2$, triggering extensive research on this material regarding the XMR origin. Since WTe$_2$ is a layered compound with metal layers sandwiched between adjacent insulating chalcogenide layers, this material has been considered to be electronically two-dimensional (2D). Here we report two new findings on WTe$_2$: (1) WTe$_2$ is electronically 3D with a mass anisotropy as low as $2$, as revealed by the 3D scaling behavior of the resistance $R(H,\theta)=R(\varepsilon_\theta H)$ with $\varepsilon_\theta =(\cos^2 \theta + \gamma^{-2}\sin^2 \theta)^{1/2}$, $\theta$ being the magnetic field angle with respect to c-axis of the crystal and $\gamma$ being the mass anisotropy; (2) the mass anisotropy $\gamma$ varies with temperature and follows the magnetoresistance behavior of the Fermi liquid state. Our results not only provide a general scaling approach for the anisotropic magnetoresistance but also are crucial for correctly understanding the electronic properties of WTe$_2$, including the origin of the remarkable 'turn-on' behavior in the resistance versus temperature curve, which has been widely observed in many materials and assumed to be a metal-insulator transition.
\end{abstract}


\maketitle

Materials exhibiting large magnetoresistances, where their resistances change significantly with applied magnetic field are a key ingredient in modern electronic devices such as hard drives in computers.\cite{ref1} The recent discovery\cite{ref2} of extremely large magnetoresistances (XMRs) in WTe$_2$ has triggered an extensive research to uncover the origin of XMRs in this material.\cite{ref2,ref3,ref4,ref5,ref6,ref7,ref8,ref9,ref10,ref11,ref12} Since it is a layered compound with metal layers sandwiched between adjacent insulating chalcogenide layers, WTe$_2$ is typically considered to be electronically two-dimensional (2D) whereby the anisotropic magnetoresistance is attributed only to the perpendicular component of the magnetic field $H\cos\theta$, where $\theta$ is the angle between the magnetic field $H$ and the crystalline c-axis.\cite{ref2,ref10} A flat band lying below the Fermi surface is ascribed\cite{ref3} to be the source of the remarkable transformative ('turn-on') temperature behavior of the sample resistance, which first decreases with temperature and then increases rapidly at low temperatures.\cite{ref2} Here, we show that WTe$_2$ is in fact electronically 3D with a small temperature dependent anisotropy. The resistance follows a 3D scaling behavior\cite{ref13} $R(H,\theta)=R(\varepsilon_\theta H)$  with $\varepsilon_\theta=(\cos^2\theta+\gamma^{-2}\sin^2\theta)^{1/2}$ and $\gamma$ being the mass anisotropy, which varies from $1.9$ to $5$.  Furthermore, we demonstrate the strong association of the anisotropy with the observed XMR where $\gamma$ follows the monotonical increase of the magnetoresistance in the Fermi liquid state when the temperature is lowered.

We measured two samples that were cleaved out of bulk crystals purchased from HQ Graphene, the Netherlands.\cite{ref14} Their thicknesses are $190$ nm (sample I) and $410$ nm (sample II), respectively. Electric contacts with well defined separations and locations were achieved using photolithography followed by evaporation deposition of $300$-$500$ nm thick Au layer with a $5$ nm thick Ti adhesion layer. DC four-probe resistive measurements were carried out in a Quantum Design PPMS-9 using a constant current mode ($I = 100$ $\mu$A). An optic image of sample I is given as Fig.S1. Angular dependencies of the resistance were obtained by placing the sample on a precision, stepper-controlled rotator with an angular resolution of $0.05^\circ$. The magnetic field is always perpendicular to the current $I$ which flows along the a-axis of the crystal. More information on the relation of the a-axis, b-axis and the direction of current flow can be found in Fig.S1.

 \begin{figure*}
 \includegraphics[width=0.80\textwidth]{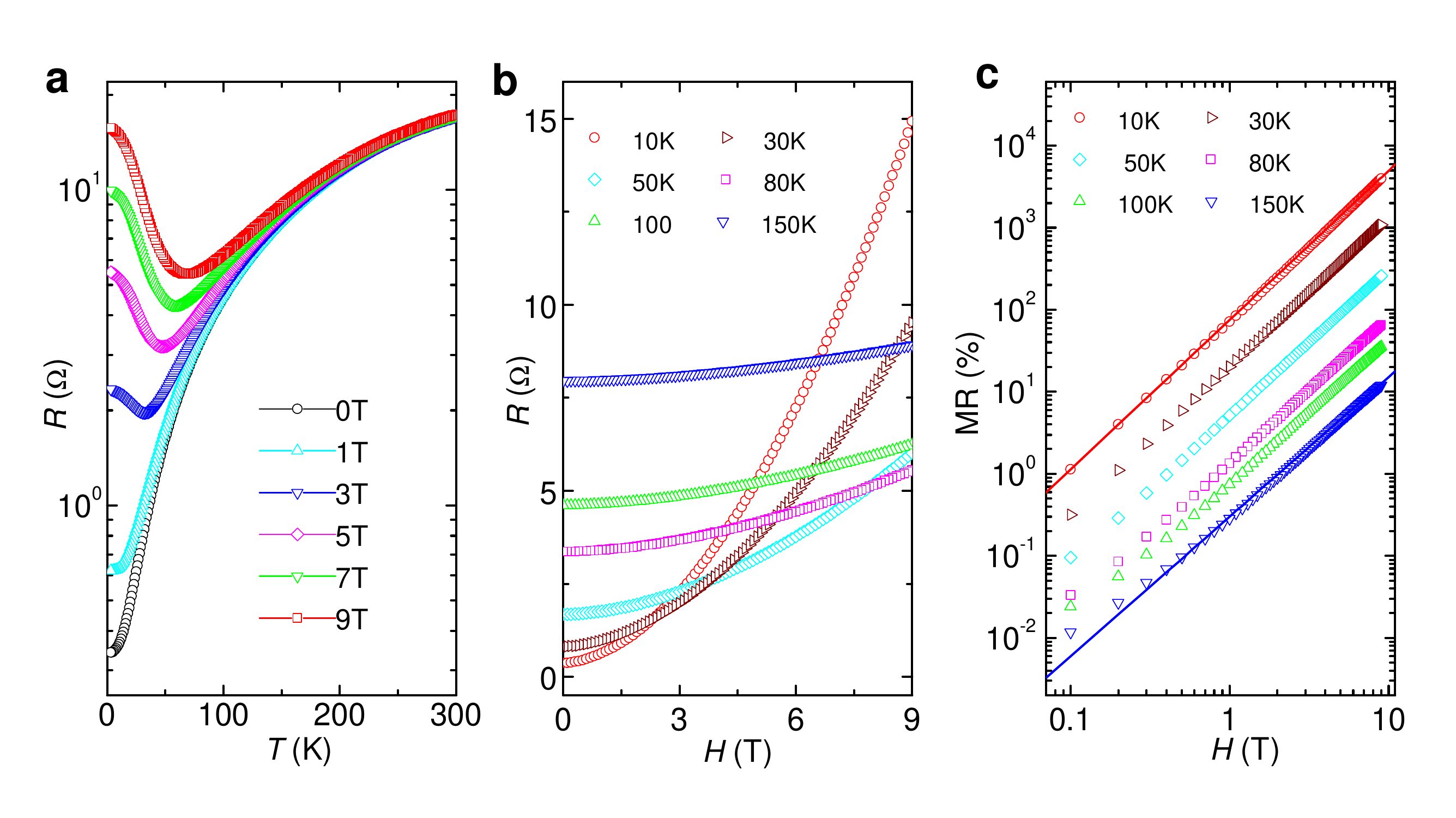}%
 \caption{\label{fig:fig1} (color online) Usual characterizations of sample I: \textbf{a}, Resistance versus temperature $R(T)$ curves at various magnetic fields. \textbf{b}, Resistance versus magnetic field $R(H)$ curves at various temperatures. \textbf{c}, Magnetoresistance versus magnetic field curves, $MR(H)$, in log-log plot, demonstrating the power relationship, where the solid lines are fits of $MR \sim H^n$ with $n  = 1.82$ at $T = 10$ K and $n = 1.7$ at $T = 150$ K. Data were taken with magnetic fields applied along c-axis of the crystal ($H\parallel c$) and we focus on the longitudinal resistance in this investigation.}
 \end{figure*}

Figure \ref{fig:fig1} presents the typical magneto-transport behavior of a WTe$_2$ crystal (sample I) at $H\parallel c$. Figure \ref{fig:fig1}a shows the temperature dependence of the resistance $R(T)$ of sample I.   In the absence of magnetic field, the resistance decreases monotonically with temperature, similar to those reported in the literature.\cite{ref2,ref4,ref8,ref10,ref12} When an external magnetic field is applied along the c-axis, the sample resistance increases and a remarkable 'turn-on' behavior appears in the $R(T)$ curves at high magnetic fields ($H \geq 3$ T), where the temperature behavior of the resistance  changes from ‘metallic’ at high temperatures to 'insulating' at low temperatures. The data also indicate that the amplitude of the magnetic field induced change in resistance increases with decreasing temperature and increasing magnetic field, consistent with the field dependence of the resistance $R(H)$ taken at various fixed temperatures presented in Fig.\ref{fig:fig1}b.  The magnetoresistance, which is defined as $MR = [R(H)-R(0)]/R(0)$ and reported as a percentage, follows the typical power-law $MR \sim H^n$ behavior with $n$ close to 2 as shown in Fig. \ref{fig:fig1}c.

\begin{figure*}
 \includegraphics[width=0.77\textwidth]{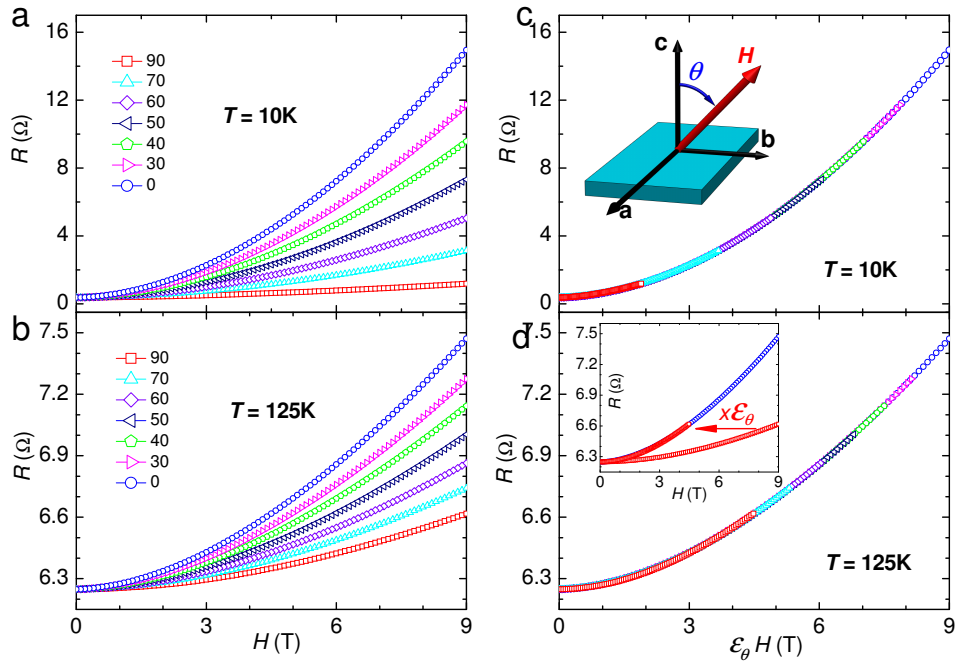}%
 \caption{\label{fig:fig2} (Color online) Scaling behavior of the field dependence of the resistances obtained at various magnetic field orientations: \textbf{a}, \textbf{b}, $R(H)$ curves of sample I at various angles $\theta$  obtained at $10$ K and $125$ K, respectively; \textbf{c}, \textbf{d}, data in \textbf{a} and \textbf{b} re-plotted with $H$ scaled by a factor $\varepsilon_\theta$. The insets of \textbf{c} and \textbf{d} show the definition of angle $\theta$ (the current flows in the a-b plane along the a-axis) and a schematic for the scaling operation (the data at $90^\circ$ and $125$ K were used as an example), respectively.}
 \end{figure*}

In order to study the anisotropy of the resistance, we measured $R(H,\theta)$ at a fixed temperature. We chose a particular applied magnetic field angle $\theta$ with respect to c-axis of the crystal (see inset of Fig.2c for the definition of $\theta$) and swept the magnetic field $H$ to obtain $R(H)$. The data obtained for sample I at $T = 10$ K and $125$ K  at various $\theta$ are presented in Figs.\ref{fig:fig2}a and \ref{fig:fig2}b, respectively. They clearly reveal that the resistance is anisotropic, with larger resistance for a fixed magnetic field applied closer to the c-axis ($\theta = 0^\circ$) for both temperatures. As discussed below in more detail, the data also indicate that the anisotropy associated with the change in resistance with $\theta$ is temperature dependent, i.e. larger at $10$ K than that at $125$ K.

\begin{figure}
 \includegraphics[width=0.42\textwidth]{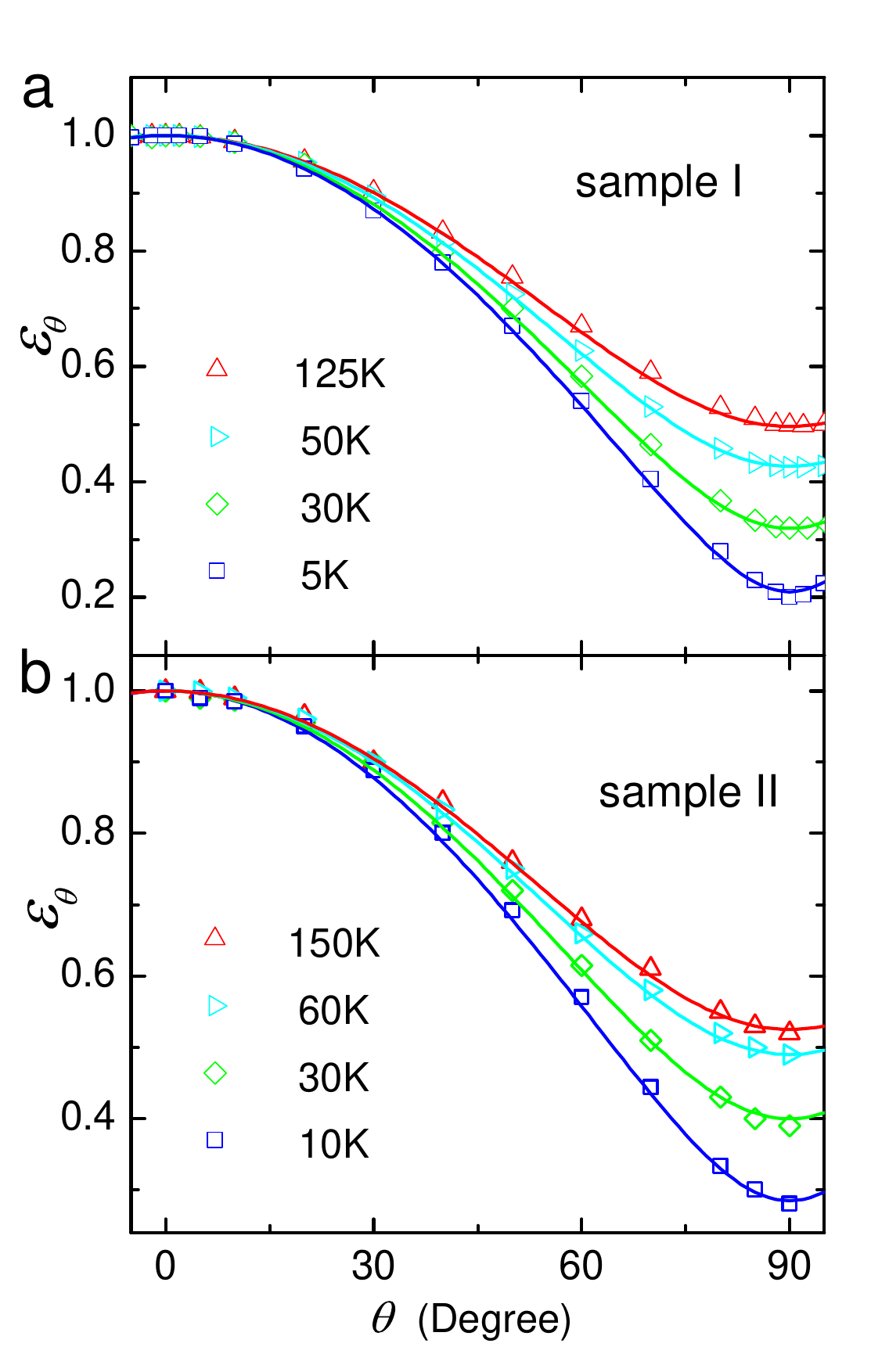}%
 \caption{\label{fig:fig3} (Color online) Angle dependence of $\varepsilon_\theta$ at various temperatures for sample I (\textbf{a}) and sample II (\textbf{b}). Symbols are derived from experimental data and lines are fits with $\varepsilon_\theta =(\cos^2\theta+\gamma^{-2}\sin^2\theta)^{1/2} $ , where $\gamma$ is a fitting parameter for a given temperature and the derived values at various temperatures for sample I are given in Fig.\ref{fig:fig4}a.}
 \end{figure}

Although the resistance anisotropy at $T = 10$ K presented in Fig.\ref{fig:fig2}a is qualitatively consistent with those reported at low temperatures in the literature (inset of Fig.\ref{fig:fig3}b in Ref.\cite{ref2}), the pronounced magnetic field dependence of the resistance at $\theta = 90^\circ$ ($H\parallel b$ axis) ($MR = 320\%$ at $10$ K and $9$ T) contradicts the expectation of a 2D system in which only the perpendicular component $H\cos\theta$ should contribute to the magnetoresistance,\cite{ref2,ref10} i.e., $R$ should be independent of $H$ at $\theta = 90^\circ$.  On the other hand, as shown in Figs.\ref{fig:fig2}c and \ref{fig:fig2}d, we found that the $R(H)$ curves obtained at a fixed temperature but at various angles can be collapsed onto a single curve, $R(H, 0^\circ)$ data, with a field scaling factor $\varepsilon_\theta =(\cos^2\theta+\gamma^{-2}\sin^2\theta)^{1/2} $  where $\gamma$ is a constant at a given temperature but changes from $4.762$ at $T = 10$ K to $2.008$ at $T = 125$ K. The temperature dependence of the scaling factor $\varepsilon_\theta$ for Sample I and a second Sample II is shown in Figs. \ref{fig:fig3}a and \ref{fig:fig3}b, and illustrates the nice agreement with the experimental data.  That is, the resistance of WTe$_2$ has the scaling behavior: 
\begin{equation}
R(H,\theta)=R(\varepsilon_\theta H)
\label{equ:equ1}
\end{equation}
with  $\varepsilon_\theta =(\cos^2\theta+\gamma^{-2}\sin^2\theta)^{1/2} $.

\begin{figure}
 \includegraphics[width=0.47\textwidth]{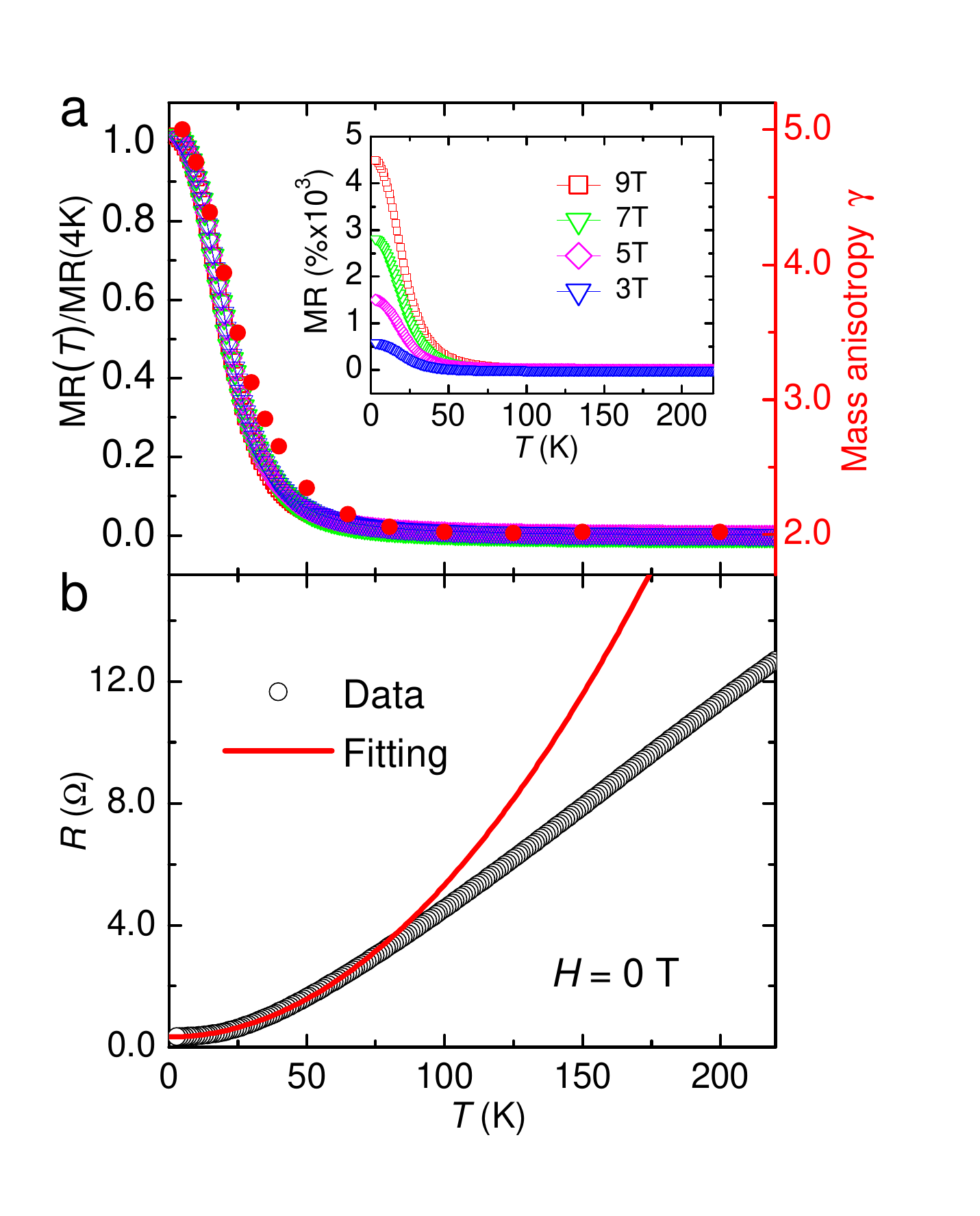}%
 \caption{\label{fig:fig4} (Color online) Association of the anisotropy $\gamma$, XMR and Fermi liquid state for sample I: \textbf{a}, temperature dependence of the $MR$ (open symbols), $\gamma$ (red solid circles); \textbf{b}, temperature dependence of zero-field resistance (open circles). In \textbf{a}, the original $MR$s and the normalized values are given in the insets and in the main panels, respectively. In \textbf{b}, symbols are the experimental data and the line represent a fit with the Fermi liquid model $R = \alpha +\beta T^2$.}
 \end{figure}

The above scaling behavior resembles that proposed for understanding the anisotropic properties of high temperature (high-T$_c$) superconductors:\cite{ref13} the angle dependence of an anisotropic quantity $Q(H,\theta)$ has the scaling behavior  $Q(H,\theta) = Q(\varepsilon_\theta H)$, where $\varepsilon_\theta H$ is the ‘reduced magnetic field’ and $\varepsilon_\theta =(\cos^2\theta+\gamma^{-2}\sin^2\theta)^{1/2} $  reflects the mass anisotropy for an elliptical Fermi surface, with $\gamma^2$ being the ratio of the effective masses of electrons moving in directions of $\theta = 0^\circ$ and $90^\circ$.  In the semi-classical model, the resistance  $R$ of a material is directly related to the mobility $\mu$ of the charge carriers through the relation $R = 1/ne\mu$ where $\mu = e\tau/m$, with $m$ being the effective mass, $\tau$ the relaxation time, and $e$ the electron charge. Thus, the anisotropy of the effective mass is expected to play a critical role in the anisotropic magnetoresistance. For example, using a two-band model, Noto and Tsuzuku\cite{ref15} theoretically obtained the angle-dependence of the magnetoresistance for graphite as $MR = A(\varepsilon_\theta H)^2/[B+C(\varepsilon_\theta H)^2]$, where $A$, $B$ and $C$ are constants and $\varepsilon_\theta =(\cos^2\theta+\alpha\sin^2\theta)^{1/2} $  with $\alpha=X^{-2}$ and $X = 12.1$ being the Fermi surface anisotropy $k_z/k_x$.\cite{ref15,ref16}  Since $X$ can also be described as $X^2 = m_\parallel/m_\perp$, where  $m_\parallel$ and $m_\perp$ are the effective masses parallel and perpendicular to the c-axis,\cite{ref17} the theory developed to account for the angle-dependence of the magnetoresistance in graphite is consistent with the general scaling rule for anisotropic superconductors and can be directly applied to understand the observed scaling behavior Eq.\ref{equ:equ1} for WTe$_2$. That is, the value of $\gamma$ obtained in our resistance scaling most likely reflects WTe$_2$'s Fermi surface anisotropy. As presented in Fig.\ref{fig:fig4}a, $\gamma$ is close to $2$ at high temperatures ( $>100$ K), which is definitely much smaller than what one would naively expect for a 2D system.\cite{ref8} It is even smaller than that (12.1) of graphite\cite{ref15,ref16} and that ($\sim 8$) of the well-known 3D high-Tc superconductor YBa$_2$Cu$_3$O$_7$.\cite{ref18} However, our results are consistent with the latest quantum oscillation experiments on WTe$_2$, which reveal a 3D Fermi surface of moderate anisotropy.\cite{ref8} The small anisotropy is probably due to the strong coupling caused by the distortion of the tellurium layers to accommodate the buckled zigzag tungsten chains.\cite{ref2,ref3}

A striking feature of the XMR in WTe$_2$ is the 'turn-on' temperature behavior: in a fixed magnetic field above a certain critical value $H_c$, a 'turn-on' temperature $T^*$ is observed in the $R(T)$ curve, where it exhibits a minimum at a field dependent temperature $T^*$. At $T < T^*$, the resistance increases rapidly with decreasing temperature while at $T > T^*$, it decreases with temperature.\cite{ref2} This 'turn-on' temperature behavior, which is also observed in many other XMR materials such as graphite,\cite{ref19,ref20} bismuth,\cite{ref20} PtSn$_4$,\cite{ref21} PdCoO$_2$,\cite{ref22} NbSb$_2$,\cite{ref23} and NbP,\cite{ref24} is commonly attributed to a magnetic-field-driven metal-insulator transition and believed to be associated with the origin of the XMR.\cite{ref10,ref19,ref20,ref23,ref25}. Khveshchenko\cite{ref25} predicted that at $T \leq T^*$ an excitonic gap $\Delta$ can be induced by a magnetic field in the linear spectrum of the Coulomb interacting quasiparticles with magnetic field dependence $\Delta(H\rightarrow H_c) \propto (H-H_c)^{1/2}$. However, the MRs obtained in our WTe$_2$ crystals at different magnetic fields as shown in Fig. \ref{fig:fig4}a, have the same temperature dependence, inconsistent with the existence of a magnetic field dependent gap, which should result in a steeper slope in the $MR$ versus $T$ curve at a higher field. Furthermore, we do not see any gap-opening induced distinctive features such as steps at $T^*$ in the $MR$ versus $T$ curves in Fig.\ref{fig:fig4}a. Thus, the observed 'turn-on' temperature behavior in WTe$_2$ is probably not due to a metal-insulator transition. In fact, thermopower experiments, which is capable of detecting a gap at the Fermi surface in the insulating state, also provide no evidence for the existence of an excitonic gap in graphite at low temperatures.\cite{ref26} On the other hand, recent angle-resolved photoemission spectroscopy (ARPES) experiments have revealed that the Fermi surfaces of WTe$_2$ differ between $20$ K and $100$ K, indicating that the observed XMR phenomenon in WTe$_2$ may be related to a change in the electronic structure with temperature. 

Our second finding, presented in the main panel of Fig.\ref{fig:fig4}a and Fig.S2a, is that the temperature dependence of $\gamma$, which is derived from the anisotropic magnetoresistance and probably directly related to the anisotropy of the Fermi surface, follows that of the $MR$s: at low temperatures ($T < 75$ K), both $MR$ and $\gamma$ increase rapidly with decreasing temperature. At $T > 100$ K where $MR$ is negligible, $\gamma$ becomes small and virtually temperature independent. That is, the XMR in our WTe$_2$ crystals is probably linked to the electronic structure, more specifically, to the Fermi surface anisotropy, which can change with temperature due to the thermal expansion of the crystal and/or temperature dependent electron-phonon coupling.\cite{ref3} As presented in Fig.\ref{fig:fig4}b, the temperature dependence of the zero-field resistance of our WTe$_2$ crystals indeed shows a transition from linear behavior originating from the electron-phonon coupling at high temperatures to the $\alpha + \beta T^2$ behavior of a Fermi liquid state with dominant electron-electron scattering at low temperatures.\cite{ref27} More importantly, the data in Fig.\ref{fig:fig4}a also clearly show that $\gamma$ starts to increase when the system enters the Fermi liquid state, enabling us to conclude that the XMR in WTe$_2$ occurs in the Fermi liquid state and its magnitude is positively correlated with the Fermi surface anisotropy. Similar results were obtained in sample II.\cite{ref28}

As evidently demonstrated in a wide range of superconductors, the mass anisotropy is the determining factor for the observed anisotropic properties.\cite{ref13,ref29} Thus, we expect that Eq.\ref{equ:equ1} can be applicable in understanding the magnetoresistance anisotropy in other materials, which is often presented as $R(\theta)$ at a particular magnetic field value $H$, i. e., the resistance is taken at a constant magnetic field while rotating the sample with respect to the external magnetic field. For example, the known Voight-Thomson relation\cite{ref15} $MR = M_1(H)\cos^2\theta + M_2(H)\sin^2\theta$ is a direct outcome of Eq.\ref{equ:equ1} if the magnetic field dependence of the resistance at $\theta = 0^\circ$ is quadratic. The $MR \sim (H\cos\theta)^{1.78}$ relation for graphite\cite{ref30} is also approximated by Eq.\ref{equ:equ1}, because at $\theta = 0^\circ$ $MR \sim H^{1.78}$ and the anisotropy $\gamma (\geq 12.1)$ is large.\cite{ref15,ref16} In fact, Eq.\ref{equ:equ1} is the only versatile way to account for $R(\theta)$, since the $MR$'s magnetic field dependence at $\theta = 0^\circ$ varies from sample to sample or even changes with temperature for the same sample. For example, the $MR(H)$ curves for most XMR materials follow a power law relationship, but the exponent ranges from $1.6$-$2.5$.\cite{ref2,ref10,ref17,ref21,ref23,ref31} In some cases, the $MR(H)$ cannot be described with a simple function.\cite{ref32} As demonstrated in Fig.S3 for our sample,\cite{ref28} in which the $MR \sim H^n$ with temperature dependent $n$, Eq.\ref{equ:equ1} can account for $R(\theta)$ curves obtained at different temperatures. 

In summary, we find that WTe$_2$ is electronically 3D with a small mass anisotropy, which varies with temperature and follows the magnetoresistance behavior of the Fermi liquid state. These findings are crucial for correctly understanding the electronic properties, including the origin of the remarkable 'turn-on' behavior in the resistance versus temperature curve, of WTe$_2$ and other XMR materials. We provide a general scaling approach for the anisotropic magnetoresistance, which is expected to account for the angle-dependence of the magnetoresistance in many other systems, where various origins have been proposed and different fitting formulas have been applied. 

\begin{acknowledgments} This work was supported by DOE BES under Contract No. DE-AC02-06CH11357 which also funds Argonne’s Center for Nanoscale Materials (CNM) and Electron Microscopy Center (EMC) where the nano
patterning and morphological analysis were performed. L.R.T. and Z.L.X. acknowledge NSF Grant No. DMR-1407175.
\end{acknowledgments}

\bibliography{AnisotropyWTe2}

\end{document}